\documentclass[onecolumn,authoryear]{els-mrw} 

\usepackage{amsmath,amssymb,amsfonts,amsthm,makeidx,graphicx}
\usepackage{txfonts}
\usepackage{helvet}
\usepackage{bm}

\usepackage{subcaption} 

\usepackage{booktabs}
\usepackage{bigstrut}
\usepackage{xspace}
\usepackage{multirow}
\usepackage{amsmath}
\usepackage{graphicx}
\usepackage{threeparttable}
\usepackage{enumitem}

\usepackage{xcolor,colortbl}
\definecolor{linkcolor}{rgb}{0.6,0,0}
\definecolor{citecolor}{rgb}{0,0,0.75}
\definecolor{urlcolor}{rgb}{0.12,0.46,0.7}
\usepackage[breaklinks=true, colorlinks, urlcolor=urlcolor, linkcolor=linkcolor,citecolor=citecolor,pdfencoding=auto]{hyperref}

\newcommand{\nver}{\hat{\mathbf{n}}}
\newcommand{\lcdm}{$\Lambda$CDM}
\newcommand{\be}{\begin{equation}}
\newcommand{\ee}{\end{equation}}
\newcommand*\diff{\mathop{}\!\mathrm{d}}

\def\aap{Astron.\ Astrophys.}

\def\apj{Astrophys.\ J.}
\def\apjl{Astrophys.\ J.\ Lett.}
\def\apjs{Astrophys.\ J.\ Suppl.\ Ser.}
\def\araa{Annual\ Review\ of\ Astron\ and\ Astrophysis.}
\def\azh{Astronomicheskii\ Zhurnal}

\def\jcap{J.\ Cosmol.\ Astropart.\ Phys.}

\def\nat{Nature}
\def\mnras{Mon.\ Not.\ R.\ Astron.\ Soc.}

\def\pasp{Publications\ of\ the\ ASP}

\def\physrep{Phys.\ Rep.}

\def\prd{Phys.\ Rev.\ D}

\def\prl{Phys.\ Rev.\ Lett.}

\begin{document}

\chapter{The Cosmic Microwave Background - Secondary Anisotropies}\label{chap1}

\author[1,2,3]{Federico Bianchini}%
\author[1,2,3]{Abhishek Maniyar}%


\address[1]{SLAC National Accelerator Laboratory, 2575 Sand Hill Road, Menlo Park, CA, 94025, USA}
\address[2]{Kavli Institute for Particle Astrophysics and Cosmology,
Stanford University, 452 Lomita Mall, Stanford, CA, 94305, USA}
\address[3]{Department of Physics, Stanford University, 382 Via Pueblo Mall, Stanford, CA, 94305, USA}

\articletag{Chapter Article tagline: update of previous edition,, reprint..}

\maketitle



\begin{glossary}[Nomenclature]
\begin{tabular}{@{}lp{34pc}@{}}
CMB & Cosmic Microwave Background\\
LSS & Large Scale Structure \\
BAO & Baryonic Acoustic Oscillations\\
\lcdm{} & Lambda Cold Dark Matter (the cosmological concordance model)\\
ISW & Integrated Sachs-Wolfe effect\\
RS & Rees-Sciama effect\\
ML & Moving Lens effect\\
(t/k)SZ & (Thermal/Kinematic) Sunyav-Zel'dovich effect\\
ACT & Atacama Cosmology Telescope\\
SPT & South Pole Telescope \\
QE & Quadratic Estimator \\
$C_L^{\phi\phi}$ & CMB lensing auto-spectrum \\
$y$ & Comptonization parameter\\
$\tau$ & Optical depth parameter
\end{tabular}
\end{glossary}

\begin{abstract}[Abstract]
The cosmic microwave background (CMB), the relic radiation from the early Universe, offers a unique window into both primordial conditions and the intervening large-scale structure (LSS) it traverses. Interactions between CMB photons and the evolving Universe imprint secondary anisotropies—modifications to the CMB’s intensity and polarization caused by gravitational effects and scattering processes. These anisotropies serve as a powerful probe of fundamental physics while also revealing astrophysical processes governing the thermodynamics and distribution of baryonic matter. In this chapter, we provide a comprehensive review of the physical mechanisms underlying secondary anisotropies, their observational status, and their potential to advance precision cosmology. With the increasing sensitivity of CMB experiments and synergy with LSS surveys, secondary anisotropies are poised to unveil unprecedented insights into the Universe’s composition, evolution, and dynamics.

\end{abstract}

\begin{BoxTypeA}[]{Key Points}
\begin{itemize}
    \item The CMB provides a snapshot of the Universe at the time of recombination, while also acting as a backlight for the large-scale structure (LSS) it traverses.
    \item Secondary anisotropies arise from two main mechanisms: gravitational effects (e.g., lensing and the Integrated Sachs-Wolfe effect) and scattering processes (e.g., Sunyaev-Zel’dovich effects and patchy screening).
    \item Secondary anisotropies are powerful tools for mapping gravitational potentials, as well as the distribution and thermal state of  gas across cosmic time.
    \item Current experiments like \textit{Planck}, ACT, and SPT, along with next-generation surveys such as the Simons Observatory and CMB-S4, promise transformative insights into secondary anisotropies, particularly through synergies with LSS surveys.
\end{itemize}
\end{BoxTypeA}

\section{Introduction}\label{chap1:intro}
The cosmic microwave background (CMB) provides our earliest view of the Universe, offering a snapshot of its state at the time of recombination and decoupling, when neutral hydrogen atoms formed. 
Observations of the CMB from full-sky satellite missions, such as WMAP \citep{Komatsu2008} in the 2000s and \textit{Planck} \citep{planck18-6} in the 2010s, have revolutionized cosmology, serving as a goldmine of information about the Universe's origins, composition, and evolution. 
These observations have played a pivotal role in establishing the concordance cosmological model, also known as the \lcdm{} model.
At the same time, high-resolution, low-noise ground-based experiments, such as the South Pole Telescope \citep[SPT,][]{Carlstrom2011,sobrin22} and the Atacama Cosmology Telescope \citep[ACT,][]{swetz11,henderson16}, have extended our knowledge of the CMB by probing smaller angular scales and polarization with unprecedented sensitivity. Upcoming experiments, including the Simons Observatory \citep[SO,][]{Ade2019}, and CMB-S4 \citep{CMB-S4_22} promise to further refine our understanding by pushing the boundaries of precision cosmology.

The CMB encodes the acoustic fluctuations imprinted in the primordial plasma, making it extremely sensitive to the Universe’s initial conditions and composition \citep[e.g.,][]{Silk67,sachs67,peebles70,doroshkevich78}. 
However, this pristine snapshot does not tell the whole story. 
The CMB also acts as a backlight for the intervening large-scale structure (LSS), and its photons interact with cosmic structures on their journey to us. 
These interactions alter the frequency, energy, and direction of propagation of the CMB photons, generating what are known as secondary anisotropies.
Secondary anisotropies are classified into two broad families based on their underlying physical mechanisms. 
The first family arises from gravitational effects, including weak gravitational lensing, decaying gravitational potentials (the integrated Sachs-Wolfe effect, or ISW), the Rees-Sciama effect (RS), and the moving lens (ML) effect.
These anisotropies are sourced by the interaction of CMB photons with gravitational potential wells. 
The second family originates from scattering processes, such as inverse Compton scattering (the Sunyaev-Zel’dovich, or SZ effects) and Thomson scattering (e.g., patchy screening). 
These effects involve interactions between CMB photons and free electrons in the intervening cosmic gas.

\begin{figure*}
    \centering
    \includegraphics[width=0.8\textwidth]{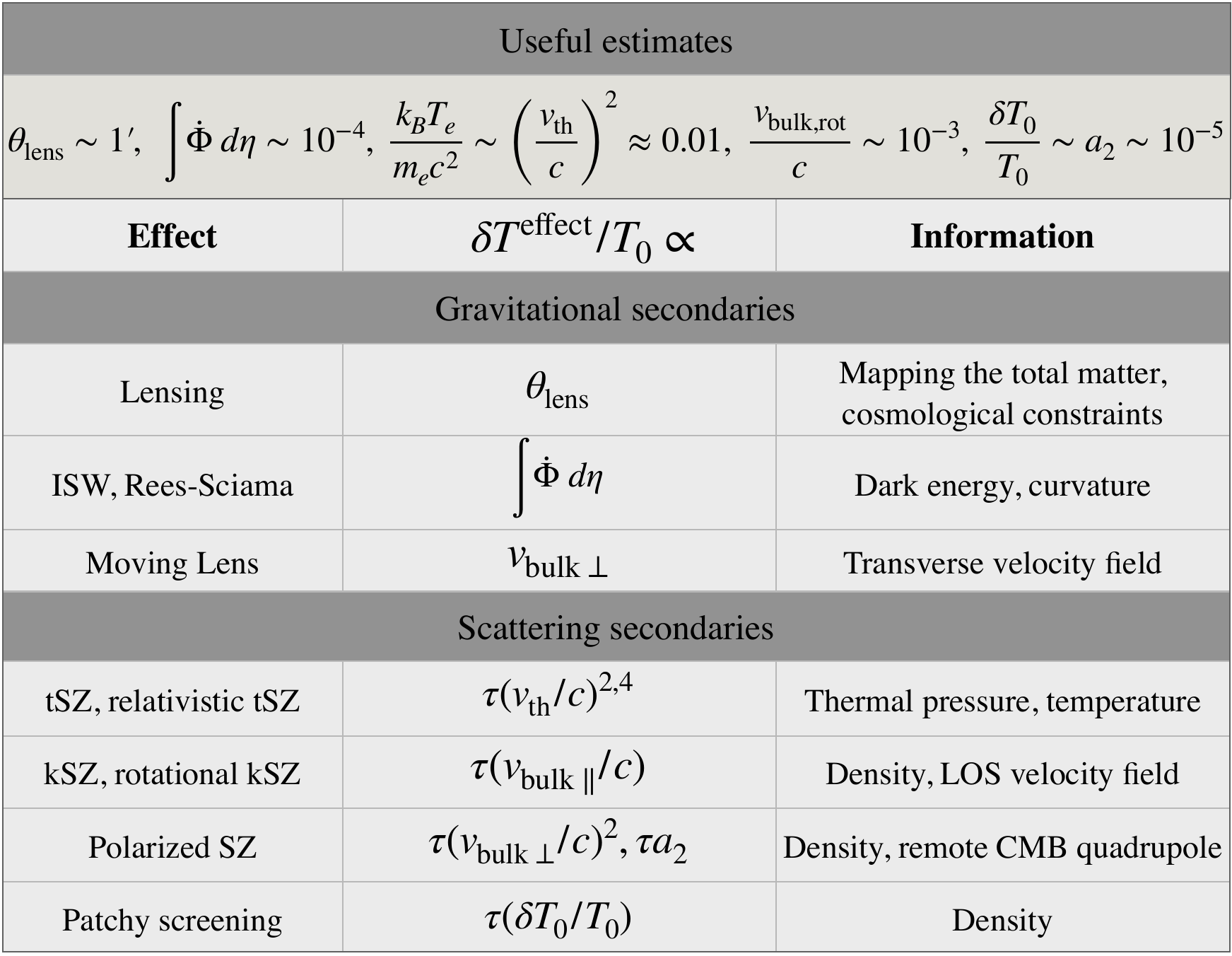}
    \caption{Summary of various secondary CMB anisotropies, highlighting their order-of-magnitude estimates and the types of information they provide about the Universe. The table distinguishes between gravitational and scattering secondaries, with key parameters such as typical lensing deflection angle ($\theta_{\rm lens}$), CMB temperature fluctuations ($\delta T_0/T_0$), and velocity terms ($v_{\rm th}$, $v_{\rm bulk}$, $v_{\rm rot}$) used to describe their effects. Symbols $\parallel$ and $\perp$ denote components parallel and perpendicular to the line-of-sight, while $a_2$ represents the CMB temperature quadrupole. Special thanks to Emmanuel Schaan for permission to reproduce this figure.}
    \label{fig:cmb_secondaries}
\end{figure*}

Secondary anisotropies encode a wealth of information about both fundamental physics—such as the mass of neutrinos, and the properties of dark matter and dark energy—and astrophysics, including the thermodynamics of gas over cosmic time. By studying these effects, summarized in Fig.~\ref{fig:cmb_secondaries}, we can extract insights into the composition and evolution of the Universe, as well as the processes shaping its structure.
This chapter is devoted to reviewing the physical mechanisms behind these secondary anisotropies and their observational status, as well as highlighting the transformative potential of upcoming surveys. 
For additional reviews on the subject, see, e.g., \citet{2002ARA&A..40..171H, Aghanim_2008}.

\section{Gravitational Secondaries}
\label{sec:grav_secondaries}
\begin{figure*}
    \centering
    \includegraphics[width=0.7\textwidth]{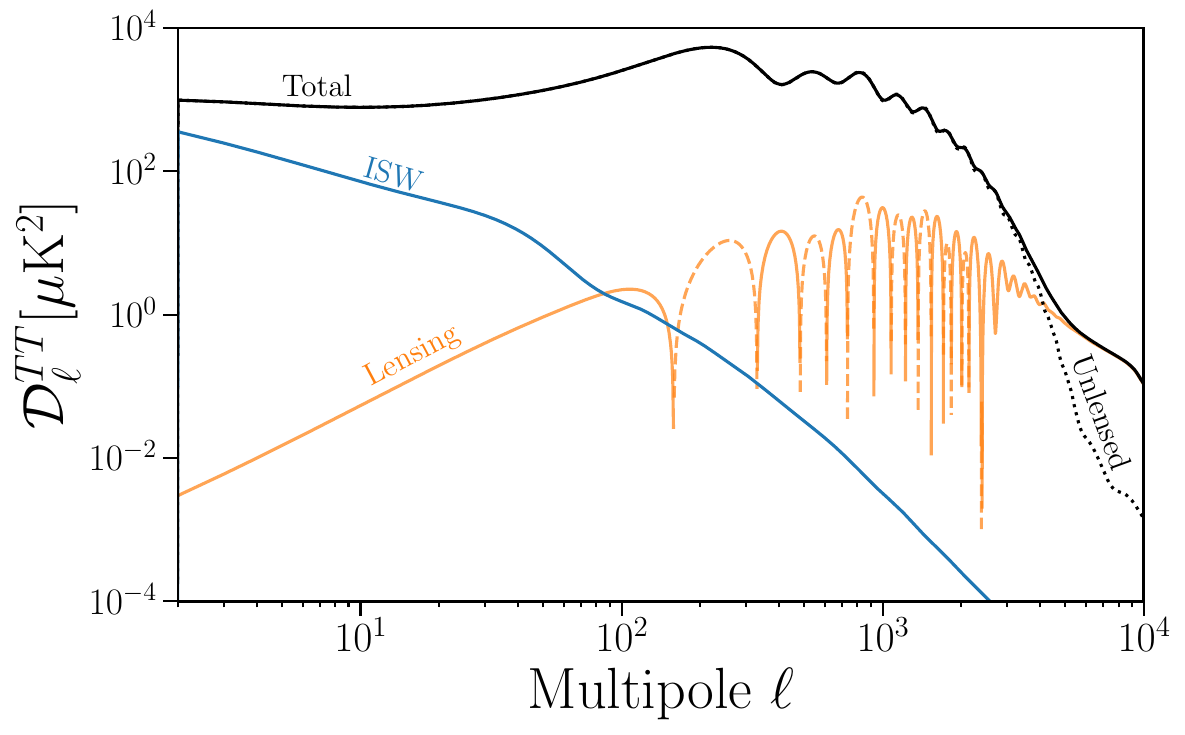}
    \caption{Power spectra of secondary temperature anisotropies from gravitational effects, normalized as $\mathcal{D}_{\ell}^{TT} \equiv \frac{\ell(\ell+1)}{2\pi}C_{\ell}^{TT}$. The blue line represents the late-time Integrated Sachs-Wolfe (ISW) effect. The orange line shows the gravitational lensing contribution, calculated as the difference between the lensed (black solid line) and unlensed (black dotted line) CMB $TT$ power spectra. The dashed orange line highlights the negative component of the lensing term.}
    \label{fig:summary_gravitational_secondaries}
\end{figure*}

Gravitational secondary anisotropies arise from the interaction of CMB photons with intervening gravitational potentials along their path. These anisotropies have two primary sources: gravitational lensing and differential redshift caused by time-varying metric perturbations.
To describe these effects, we begin with the line element $\diff s^2$ of a perturbed, spatially flat Friedmann-Lemaître-Robertson-Walker (FLRW) metric. In the conformal Newtonian (or longitudinal) gauge, the metric takes the form \citep{Ma_1995}:
\begin{equation}
\label{eq:pertFLRW_conf}
\diff s^2 = a^2(\eta)\left[-(1 + 2\Psi_N)\diff\eta^2 + (1 + 2\Phi_N)\gamma_{ij}\diff x^i \diff x^j\right],
\end{equation}
where $a(\eta)$ is the scale factor depending on conformal time $\eta$, $\Phi_N$ and $\Psi_N$ are the scalar metric potentials\footnote{Here, we consider only scalar perturbations.} \citep{bardeen80}, and $\gamma_{ij}$ is the unperturbed spatial metric:
\begin{equation} 
\gamma_{ij}\diff x^i \diff x^j = \diff \chi^2 + \chi^2\left(\diff \theta^2 + \sin^2{\theta}\diff \phi^2\right), 
\end{equation} 
where $\chi$ is the comoving radial distance.
The metric in Eq.~\ref{eq:pertFLRW_conf} resembles the weak-field limit of General Relativity near Minkowski space. 
In this limit, $\Phi_N$ acts as the gravitational potential governing the dynamics of non-relativistic objects, while the Weyl potential, defined as $\Psi_{\gamma} \equiv (\Psi_N - \Phi_N)/2$, determines the propagation of light by describing null geodesics.
In the absence of anisotropic stress—associated with the trace-free component of the stress-energy tensor $T_{ij}$—the scalar potentials satisfy $\Psi_N = -\Phi_N$. 
This approximation holds in standard cosmological models during the matter- and dark-energy-dominated epochs, where contributions from anisotropic stress are negligible.
By adopting the perturbed FLRW metric in the conformal Newtonian gauge, we establish the framework for analyzing gravitational secondary anisotropies. This formalism underpins the study of effects such as gravitational weak lensing, the integrated Sachs-Wolfe (ISW) effect, the Rees-Sciama effect, and moving lens effects. 
The temperature power spectra of selected gravitational secondary anisotropies are illustrated in Fig.~\ref{fig:summary_gravitational_secondaries}.

\subsection{Gravitational weak lensing}
\label{sec:lensing}
After recombination at redshift $z \sim 1090$, CMB photons traverse an evolving and clumpy universe before reaching our telescopes. 
Along the way, their trajectories are deflected by the gravitational potentials of LSS \citep{Blanchard1987}. 
These deflections, typically on the order of 2 arcminutes and coherent over degree scales, leave distinct imprints on the observed CMB: they blur the sharp acoustic peaks \citep{seljak96b}, transfer power from large to small scales \citep{Metcalf:1997ih}, induce non-Gaussian features in the anisotropies \citep{bernardeau97b,zaldarriaga00}, and convert a fraction of $E$-mode polarization into $B$-modes \citep{zaldarriaga98}.
CMB lensing serves a dual role in cosmology. 
On the one hand, it provides a valuable probe of the integrated gravitational potentials along the line-of-sight, offering insights into the distribution of matter and the growth of cosmic structures \citep[e.g.,][]{Stompor_1999,Acquaviva2006}. 
On the other hand, it acts as an obstacle by obscuring our view of primordial fluctuations, complicating the measurement of the tensor-to-scalar ratio parameter $r$ \citep{knox02}.
The CMB is particularly advantageous for lensing studies due to its unique properties. 
As a single, well-defined source plane at $z \sim 1090$, the CMB lies behind all structure in the observable Universe.
This unique position allows it to probe the LSS over the entire sky, complementing galaxy lensing observations, which typically focus on lower redshifts. Unlike galaxy surveys, the precisely known redshift of the CMB eliminates uncertainties related to source distances. Additionally, the intrinsic fluctuations in the CMB are predominantly Gaussian, simplifying the lensing reconstruction process.
CMB lensing primarily originates from linear and mildly non-linear scales, which makes them easier to model and interpret compared to galaxy lensing, where contributions from highly non-linear scales are significant \citep[e.g.,][]{Amon_2022}. 
These properties make CMB lensing a powerful cosmological tool, enabling precise mapping of the matter distribution, constraints on key cosmological parameters, and investigations into the nature of dark energy and dark matter.
For comprehensive reviews of CMB lensing, we refer the reader to \citet{Lewis2006,Hanson_2010}.

The lensing effect on CMB anisotropies can be described as a remapping of the unlensed CMB anisotropies, $\tilde{X}(\nver)$, by the deflection angle $\mathbf{d}(\nver)$ \citep[e.g.,][]{lewis06}:\footnote{Although not an exact remapping, this approximation is sufficiently accurate for most applications \citep{Lewis_2017}.}
\begin{equation}
\label{eq:lens_effect}
X(\nver) = \tilde{X}[\nver + \mathbf{d}(\nver)],
\end{equation}
where $X(\nver)$ represents the observed temperature $T(\nver)$ or polarization $P_\pm(\nver) = [Q \pm iU](\nver)$ anisotropies. 
Lensing preserves the blackbody spectrum and surface brightness of the CMB by altering the angle and magnification of the source without changing the photon frequency or density per unit solid angle.
The deflection field $\mathbf{d}(\nver)$ in Eq.~\ref{eq:lens_effect} can be decomposed into gradient and curl components \citep[e.g.,][]{hirata03b}:
\begin{equation}
\mathbf{d}(\nver) = \boldsymbol{\nabla} \phi(\nver) + (\star \boldsymbol{\nabla}) \omega(\nver),
\end{equation}
where $\phi(\nver)$ is the scalar lensing potential, $\omega(\nver)$ is the curl component (or “pseudo-scalar lensing potential”), and $\star \boldsymbol{\nabla}$ denotes the derivative with a $90^\circ$ counterclockwise rotation in the plane perpendicular to the line-of-sight \citep{namikawa12}.

The scalar lensing potential, $\phi(\nver)$, is given by a line-of-sight integral:
\begin{equation}
\phi(\nver) = -2 \int_0^{\chi_*} \diff\chi \left(\frac{\chi_* - \chi}{\chi_* \chi}\right) \Psi_\gamma(\chi \nver, \eta_0 - \chi),
\end{equation}
where $\eta_0 - \chi$ is the conformal time when the photon was at location $\chi \nver$, $\Psi_\gamma$ is the Weyl potential (see Sec.~\ref{sec:grav_secondaries}), and the integral is evaluated along the unperturbed photon path under the Born approximation \citep[e.g.,][]{hirata03b}.
The lensing convergence, $\kappa(\nver)$, quantifies the local magnification effect on CMB anisotropies and is defined as the two-dimensional Laplacian of the lensing potential: $\kappa(\nver) = -\frac{1}{2} \nabla^2 \phi(\nver)$.
Using the cosmological Poisson equation, $\nabla^2 \Phi \propto \delta_{\rm m}$,\footnote{This assumes General Relativity and neglects minor effects from massive neutrinos.} the lensing convergence can be related to the projected matter distribution $\delta_m$ along the line-of-sight:
\begin{equation}
\kappa(\nver) = \int \diff\chi \, W^{\kappa}(\chi) \delta_m(\chi\nver, \eta_0-\chi),
\end{equation}
where the lensing kernel $W^{\kappa}(\chi)$ describes the lensing efficiency, peaking around $z \approx 2$.
At linear order, scalar metric perturbations, such as matter density fluctuations, produce only the gradient mode $\nabla \phi(\nver)$, while the curl mode is generally negligible and is often used as a systematic check in lensing analyses \citep[e.g.,][]{cooray05}. 
However, curl modes can arise from vector and tensor perturbations \citep[e.g.,][]{Li_2006,namikawa12} or higher-order effects beyond the Born approximation \citep[e.g.,][]{Pratten_2016,robertson2023}.
The Born approximation, extensively tested in simulations \citep[e.g.,][]{Calabrese15,beck18,Fabbian2018}, introduces only minor corrections. 
Post-Born corrections to the convergence power spectrum are small, at approximately $0.2\%$ \citep{Krause_2010, Pratten_2016}, and their effects on CMB temperature and polarization anisotropies have been studied in detail \citep{Hagstotz2015,Marozzi2016,Pratten_2016}.

To estimate the typical scale of the lensing effect, consider a gravitational potential depth of $|\Phi| \sim 2 \times 10^{-5}$. 
This corresponds to deflection angles of $|\mathbf{d}| \sim 10^{-4}$ radians and structure sizes of approximately 300 Mpc, as set by the scale of the peak of the matter power spectrum. 
A photon traveling from the last-scattering surface, roughly 14,000 Mpc away, encounters about $14000/300 \sim 50$ independent deflections. The cumulative root-mean-square (rms) deflection is therefore $\sqrt{50} \times 10^{-4} \sim 2$ arcminutes. 
The coherence scale of the deflection field is determined by the angular size of a typical 300 Mpc structure. 
At a comoving distance of $\chi = 7000$ Mpc (halfway to the last-scattering surface), this corresponds to an angular scale of approximately $300/7000 \sim 2$ degrees.

Lensing modifies the statistical properties of the observed CMB anisotropies. To gain intuition into its effects, we can Taylor-expand the remapping equation (Eq.~\ref{eq:lens_effect}):\footnote{This expansion is not accurate on all scales, especially for lensing at scales comparable to the deflection angles. See \citet{challinor05} for an exact treatment.}
\begin{equation}
\label{eq:lens_firs_ord_exp}
X(\nver) = \tilde{X}(\nver) + [\nabla \phi(\nver) + (\star \nabla)\omega(\nver)] \cdot \nabla \tilde{X} + \mathcal{O}(\phi^2, \omega^2),
\end{equation}
where the additional terms are proportional to the gradients of the unlensed CMB fields.
In the small-angle approximation, focusing on the temperature field and considering only the scalar potential, the first-order expansion in harmonic space becomes \citep{hu00b}:
\be
\label{eq:cmblens_pert}
T_{\bm{\ell}} \approx \tilde{T}_{\bm{\ell}} - \int \frac{\diff^2\bm{L}}{2\pi} \left[\bm{L}\cdot (\bm{\ell}-\bm{L})\right]\phi_{\bm{\ell}-\bm{L}}\tilde{T}_{\bm{L}}.
\ee
Assuming that both $\tilde{T}_{\bm{\ell}}$ and $\phi_{\bm{\ell}}$ are isotropic Gaussian random fields and uncorrelated,\footnote{Correlations such as $\langle T\phi\rangle$ induced by secondary anisotropies (e.g., the ISW and SZ effects) are neglected. See Sec.~\ref{sec:isw_rs} and Sec.~\ref{sec:sz}.} the lensed temperature power spectrum is given by:
\be
C_{\ell}^{TT} \approx \left(1-\ell^2 \int\frac{\diff\ln L}{4\pi} L^4 C_L^{\phi\phi}\right)C_{\ell}^{\tilde{T}\tilde{T}} + \int\frac{\diff^2\bm{L}}{2\pi} [\bm{L}\cdot (\bm{\ell}-\bm{L})]^2 C_{|\bm{\ell}-\bm{L}|}^{\phi\phi}C_{L}^{\tilde{T}\tilde{T}},
\ee
This equation shows that the lensed power spectrum is a convolution of the unlensed temperature gradient power spectrum $C_{L}^{\tilde{T}\tilde{T}}$ with the deflection power spectrum $C_L^{\phi\phi}$. 
As shown by the oscillations and the asymptotic positive offset in the orange line of Fig.~\ref{fig:summary_gravitational_secondaries}, lensing smooths the acoustic peaks in the CMB temperature anisotropies and enhances the small-scale power in the damping tail.
Despite reshuffling power across scales, lensing conserves the total temperature variance since it preserves brightness while altering photon directions. 
Lensing has similar effects on $C_\ell^{EE}$ and $C_\ell^{TE}$ as it does on temperature anisotropies. 
Additionally, lensing converts a fraction of $E$-mode power into $B$-modes, creating a lensed $C_\ell^{BB}$ signal \citep[e.g.][]{zaldarriaga98}. 
This leakage corresponds to an effective white-noise level of $\sim 5 \mu$K-arcmin at $\ell \lesssim 100$, posing a significant contamination to searches for primordial gravitational waves using $B$-modes on large angular scales.

Lensing also introduces mode coupling, correlating previously independent harmonic modes $\bm{\ell} \ne \bm{\ell'}$, as shown in Eq.~\ref{eq:cmblens_pert}. 
For a fixed realization of the lenses, the CMB covariance matrix acquires off-diagonal elements:
\be
\langle X_{\bm{\ell}} Y^*_{\bm{L}-\bm{\ell}}\rangle_{\rm CMB} = f^{XY}_{\bm{\ell},\bm{L}}\phi_{\bm{L}} + \mathcal{O}(\phi^2), \quad ({\bm L} \ne 0)
\ee
where $\langle\cdot\rangle_{\rm CMB}$ denotes an average over CMB realizations, and $f^{XY}_{\bm{\ell},\bm{L}}$ are coupling kernels that depend on the observable $X,Y \in [T,E,B]$. 
These mode couplings provide the foundation for constructing quadratic estimators (QEs) of the lensing potential:
\be
\hat{\phi}^{XY}_{\bm{L}} = A^{XY}_{\bm{L}}\int \frac{\diff\bm{\ell}^2}{2\pi} F^{XY}_{\bm{\ell},\bm{L}}X_{\bm{L}}Y_{\bm{\ell}-\bm{L}},
\ee
where $A^{XY}_{\bm{L}}$ normalizes the estimator to ensure unbiasedness, and $F^{XY}_{\bm{\ell},\bm{L}}$ is a filter optimized for maximal signal-to-noise ratio.
For full expressions, see \citet[e.g.,][]{Hu2002,Maniyar_2021}, and for the ``shear-only" QE, see \citet{Schaan_2019}.
Beyond QEs, Bayesian methods leveraging the full CMB posterior have been proposed to extract higher-order information and approach optimality \citep[e.g.,][]{hirata03b,Carron_2017,Millea_2020,Millea_2022}.

Once the lensing potential map is reconstructed, the lensing power spectrum can be calculated by subtracting noise bias terms \citep[e.g.,][]{Hanson_2011,Shen_2024}. 
The resulting spectrum is then compared with theoretical predictions to infer cosmological constraints. 
Within the base \lcdm{} model, CMB lensing constrains a narrow elongated tube in the 3D subspace spanned by $\sigma_8$-$H_0$-$\Omega_{\rm m}$, with the strongest constraint on the combination $\sigma_8\Omega_m^{0.25}$.
Combining CMB lensing with low-redshift galaxy BAO measurements breaks degeneracies among $\sigma_8$, $H_0$, and $\Omega_{\rm m}$, sharpening constraints on individual parameters.\footnote{See \citet{pan14,planck15-15,madhavacheril23} for detailed discussions on parameter dependencies of the CMB lensing power spectrum.} 

\begin{figure*}
    \centering
    \includegraphics[width=0.8\textwidth]{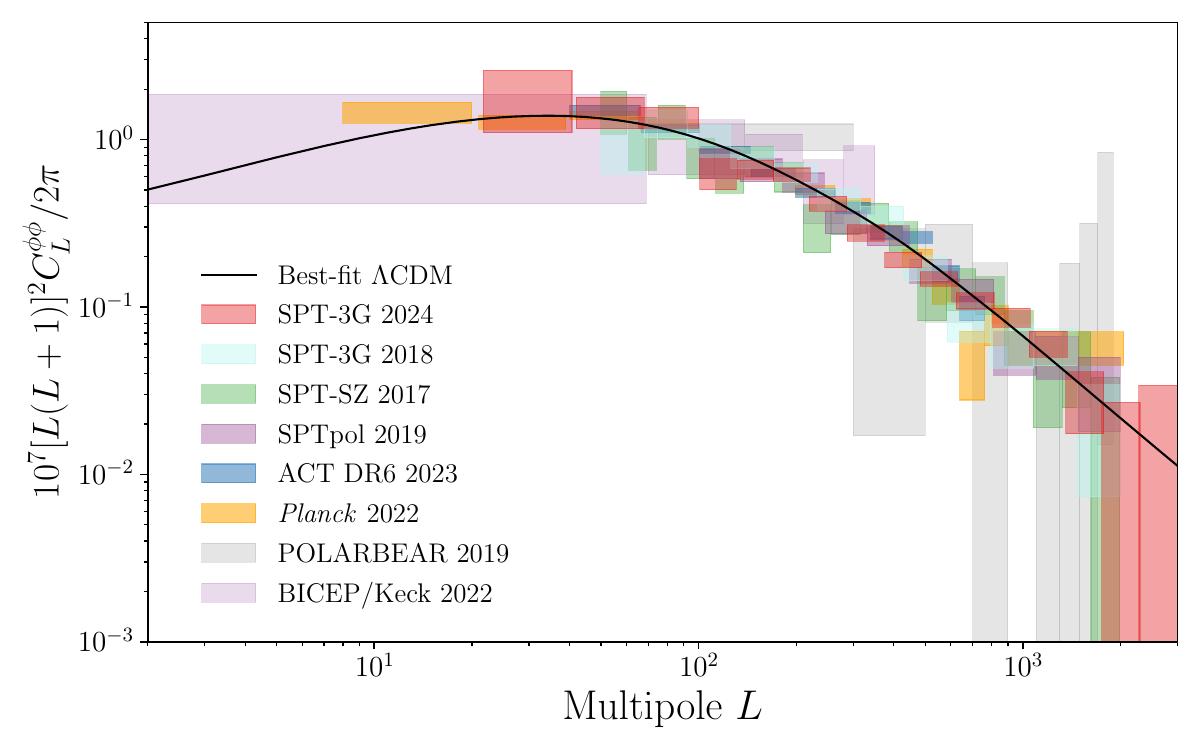}
    \caption{Compilation of CMB lensing power spectrum measurements from various ground- and space-based CMB surveys. The solid black line corresponds to the best-fit \lcdm{} model to the 2018 {\it Planck} \texttt{TT,TE,EE+lowE+lensing} dataset \citep{planck18-6}.}
    \label{fig:cmb_lensing_measurements}
\end{figure*}

Since its first detection in 2007 by cross-correlating lensing maps from the Wilkinson Microwave Anisotropy Probe (WMAP) with NRAO VLA Sky Survey radio galaxies \citep{Smith:2007rg}, CMB lensing science has advanced significantly. The \textit{Planck} team reconstructed an almost full-sky lensing map and measured the lensing power spectrum with a significance of $25\sigma$ using temperature data \citep{planck13-17}. 
More recently, high-significance measurements of the CMB lensing auto-spectrum have been achieved by ACT \citep{Das_2011,sherwin17,qu23}, SPT \citep{van_Engelen_2012,Story_2015,omori17,wu19,Pan_2023,ge2024cosmologycmblensingdelensed}, and \textit{Planck} \citep{planck13-17,planck15-15,Carron_2022}, with detection significances up to around $40\sigma$. 
A compilation of these measurements is shown in Fig.~\ref{fig:cmb_lensing_measurements}.

Cross-correlations between CMB lensing and other tracers of structure allow for the extraction of redshift-localized information about the growth rate of structure in the Universe—a technique often referred to as ``CMB lensing tomography." These correlations also provide valuable insights into astrophysical processes and offer greater control over systematic uncertainties \citep[e.g.,][]{schaan17}.
CMB lensing maps have been cross-correlated with various tracers, including galaxy catalogs \citep[e.g.,][]{sherwin12,Bianchini_2015,Peacock_2018,Bianchini_2018,Omori18,Marques:2019aug,Krolewski_2020,Darwish_2020,Alonso:2023guh}, galaxy lensing \citep[e.g.,][]{Hand:2013xua,Liu:2015xfa,Singh:2016xey,POLARBEAR:2019phb,Robertson:2020xom}, cosmic infrared background (CIB) fluctuations \citep[e.g.,][]{holder13,planck13-18,ACT:2014swt}, thermal Sunyaev-Zel’dovich Compton-$y$ maps \citep[e.g.,][]{Hill_2014,McCarthy:2023cwg}, peculiar velocity catalogs \citep[e.g.,][]{Giani:2023vfr}, and radio background maps \citep[e.g.,][]{Todarello2024}.
The complementarity of these probes, combined with their potential to break parameter degeneracies, has driven interest in joint analyses of auto- and cross-correlation spectra between multiple cosmic fields—a strategy commonly referred to as ``$N\times 2$ analyses" \citep[e.g.,][]{Nicola:2016eua,abbott19,Chang_2023,Fang:2023efj,Sgier:2021bzf}.

Finally, an exciting possibility that has emerged in recent years is to use CMB lensing to measure the mass of samples of galaxies and galaxy clusters \citep[e.g.][]{baxter15,madhavacheril15,Geach:2017crt,raghunathan17b,raghunathan19a,raghunathan19b,SPT-3G:2024lri}.
This is an especially promising technique to measure the mass and calibrate the mass-observable scaling relations of clusters at $z \gtrsim 1$, where the rate of lensed background galaxies observed with high S/N drops significantly.

\subsection{Integrated Sachs-Wolfe \& Rees-Sciama effects}
\label{sec:isw_rs}
The Integrated Sachs-Wolfe (ISW) effect is a gravitational contribution to the observed CMB temperature fluctuations, arising from the time evolution of gravitational potentials as photons travel from the surface of last scattering to us. 
Depending on the regime and the underlying physical mechanism driving the potential evolution, this effect is sometimes referred to as the Rees-Sciama (RS) effect.

By solving the geodesic equation, photons traversing a time-varying gravitational potential $\Phi_N$ experience an energy change, expressed in terms of temperature fluctuations as \citep[e.g.,][]{sachs67}:\footnote{Assuming negligible anisotropic stress, such that $\Phi_N = -\Psi_N$.}
\be
\frac{\Delta T_{\rm ISW}(\nver)}{T_{\rm CMB}}=  2\int_{\eta_0}^{\eta_*} \diff\eta e^{-\tau(\eta)}\dot{\Phi}_N(\nver, \eta),
\ee
where $\tau(\eta)$ is the optical depth, and the dot denotes differentiation with respect to conformal time $\eta$. 
This temperature shift represents the redshifting or blueshifting of photons climbing out of a potential well that has changed since they entered it.

The late-time ISW effect arises from the decay of gravitational potentials at low redshifts, driven by the accelerated expansion of the universe due to dark energy. 
The gravitational potential $\Phi_N$ can be related to the matter density fluctuations $\delta_{\rm m}$ via the Poisson equation. 
In Fourier space, under the Newtonian limit for subhorizon scales ($k \gg aH$), this relation is:
\be
\Phi_N(\mathbf{k},\eta) = \frac{3}{2}\frac{\Omega_m}{a} \left(\frac{H_0}{k}\right)^2 \delta_{\rm m}(\mathbf{k},\eta).
\ee
During matter domination, $\delta_{\rm m} \propto a$, leading to $\dot{\Phi}_N \simeq 0$ and a negligible ISW effect. 
However, at lower redshifts, when dark energy dominates, $\delta_{\rm m}$ grows more slowly than $a$, causing $\Phi_N$ to decay and $\dot{\Phi}_N \neq 0$. 
The ISW effect is achromatic (frequency-independent) and primarily contributes to large angular scales (low $\ell$) in the CMB temperature power spectrum, as shown by the blue line in Fig.~\ref{fig:summary_gravitational_secondaries}.

The strength of the ISW effect depends on the properties of dark energy (e.g., its equation of state and clustering), curvature, and potential modifications to gravity. 
However, the ISW signal is inherently weak and limited by cosmic variance. 
\citet{Crittenden1996} proposed isolating the ISW effect through cross-correlation between CMB temperature maps and large-scale structure (LSS) tracers, as both probe the same gravitational potentials. 
The first detection of the ISW was achieved by cross-correlating WMAP first-year data with radio galaxy counts from NVSS and X-ray sources from HEAO1 A1 \citep{Boughn2004}. 
Subsequent analyses with various LSS tracers have yielded ISW detections with significances up to $\sim 4\sigma$, though some results remain inconclusive due to foreground contamination and differences in statistical methodologies \citep[e.g.,][]{fosalba03,Giannantonio_2006,Granett2008,Ferraro2015,planck_isw,Krolewski2022J}. 
Cosmic voids have also been used as tracers of gravitational potentials for ISW studies \citep[e.g.,][]{Ilic_2013,DES:2018nlb}.

On smaller scales, the RS effect becomes significant, arising from non-linear gravitational collapse. 
If the photon crossing time through a structure is comparable to the potential evolution timescale, the net energy shift from the red- and blueshifting processes is nonzero, leaving an imprint on the CMB. 
Originally described by \citet{Rees:1968zza}, the RS effect dominates the ISW effect at angular scales corresponding to $\ell > 200$. 
Unlike the ISW effect, the RS effect does not require dark energy, as non-linear gravitational collapse causes $\delta_{\rm m}$ to grow faster than $a$, leading to $\dot{\Phi}_N < 0$ \citep{PhysRevD.65.083518}. 
At late times, linear scales typically exhibit $\dot{\Phi}_N > 0$, while non-linear scales exhibit $\dot{\Phi}_N < 0$, resulting in opposite correlations between tracers and the CMB \citep{Nishizawa_2008}.

The RS effect is currently too small to be detected by existing experiments.
However, futuristic CMB experiments such as CMB-S4 or CMB-HD, in combination with Rubin LSST galaxies, may achieve modest detections with signal-to-noise ratios of $S/N \sim 6-8$ \citep{Ferraro2022}.
For a detailed review of the ISW and RS effects, we refer the reader to \citet{Nishizawa2014}.

\subsection{Moving lens effect}
\label{sec:moving_lens}
The moving lens (ML) effect, also known as the "slingshot effect" \citep[e.g.,][]{Birkinshaw1983,Tuluie1995,Stebbins:2006tv}, is deeply related to the Rees-Sciama (RS) effect and arises from the transverse motion of gravitational potentials across the observer’s line-of-sight.
This motion generates a dipole signal in the CMB, aligned with the direction of motion.
Specifically, CMB photons entering ahead of a moving structure (e.g., a galaxy cluster or supercluster) are redshifted, while those in the structure’s wake are blueshifted. 
The resulting dipolar imprint is proportional to the transverse velocity $\mathbf{v}_\perp$ and the depth of the potential well, which scales with the total mass $M_{\rm tot}$ of the halo. 

For a gravitational potential moving with a peculiar (comoving) velocity $\mathbf{v}_\perp$ transverse to the line-of-sight direction $\nver$, the induced CMB temperature fluctuations are given by:
\be
\frac{\Delta T_{\rm ML}(\nver)}{T_{\rm ML}}= -2\int \diff \chi \mathbf{v}_\perp \cdot \nabla_\perp\Phi_N(\chi\nver),
\ee
where $\nabla_\perp \Phi_N$ is the gradient on the 2-sphere of the gravitational potential.
Like the ISW and lensing effects, the ML effect has a blackbody frequency dependence and cannot be isolated through multi-frequency observations.

The amplitude of the ML signal is small and subdominant compared to other signals and residual foregrounds, making its detection challenging. 
Consequently, the ML effect is more likely to be detected statistically across an ensemble of objects rather than for individual halos. 
Recent years have seen renewed interest in this effect, with the development of novel estimators, including pairwise transverse-velocity estimators, oriented stacking techniques, and quadratic estimators (QEs), to evaluate its detectability with upcoming surveys \citep{Yasini:2018rrl,Hotinli_2019,Hotinli_2021}.
For instance, \citet{hotinli2024detectabilitymovinglenssignal} suggest that next-generation Stage-4 CMB and LSS surveys could detect the ML effect with a significance of 10–20$\sigma$. 
However, ongoing Stage-3 surveys, such as DESI and SO, are expected to achieve lower detection significance due to the smaller number of observed halos.


\section{Scattering secondaries}
\begin{figure*}
    \centering
    \includegraphics[width=0.7\textwidth]{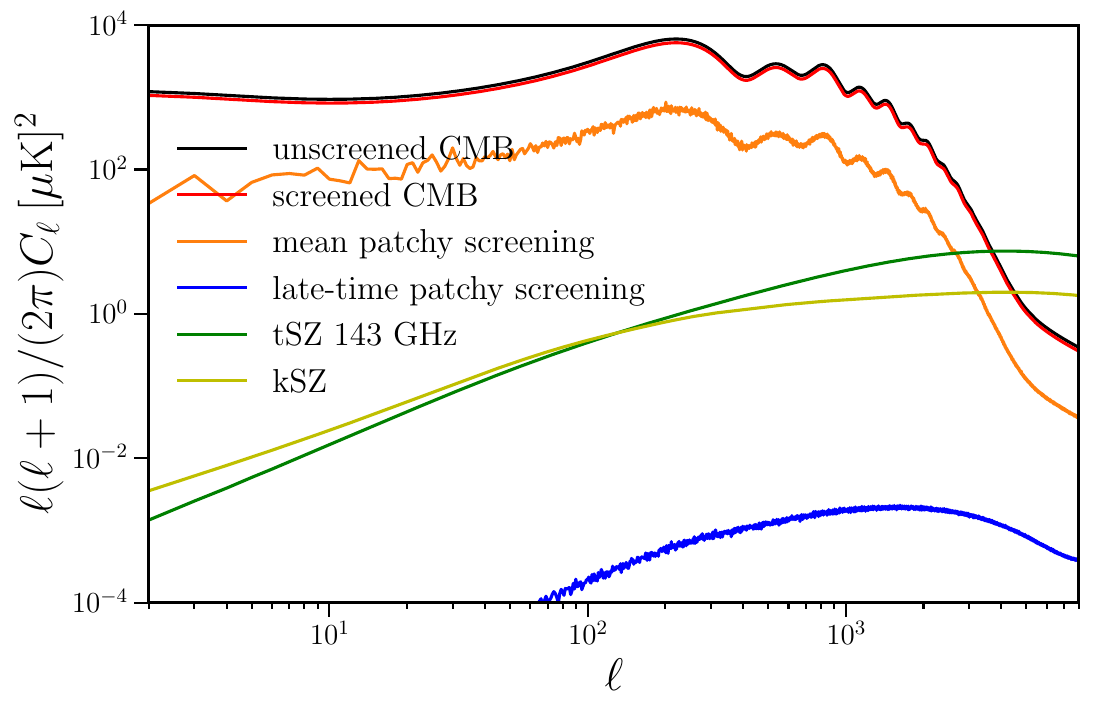}
    \caption{Power spectra of secondary temperature anisotropies from scattering effects. The green and yellow lines represent the tSZ and kSZ effects respectively. The tSZ curve has been calculated assuming bandpass filters from the \textit{Planck} experiment at 143 GHz. The orange line shows the mean patchy screening effect on the CMB power spectrum and the red curve is the resulting CMB power spectrum, calculated as the difference between the lensed (black solid line) CMB $TT$ power spectra and orange curve. The blue curve presents the contribution of the ``late-time" patchy screening ($z < 5$) calculated using a template from AGORA simulations \citep{Omori2024}.}
    \label{fig:summary_scattering_secondaries}
\end{figure*}

Scattering secondary anisotropies arise from interactions between CMB photons and free electrons in the ionized medium, altering the energy, direction, or polarization of photons. 
These anisotropies are primarily sourced by two processes: inverse Compton scattering, which leads to the thermal and kinematic Sunyaev-Zel’dovich (SZ) effects, and Thomson scattering, which produces anisotropies associated with patchy screening. 
Scattering-induced effects provide insights into the thermodynamic state and distribution of baryonic matter, as well as the history of reionization and the evolution of cosmic structures.
In Fig.~\ref{fig:summary_scattering_secondaries}, we illustrate the contributions of the SZ effects and patchy screening to the CMB temperature power spectrum.

\subsection{Sunyaev-Zel'dovich (SZ) effects}
\label{sec:sz}
\begin{figure*}
    \centering
    \includegraphics[width=0.9\textwidth]{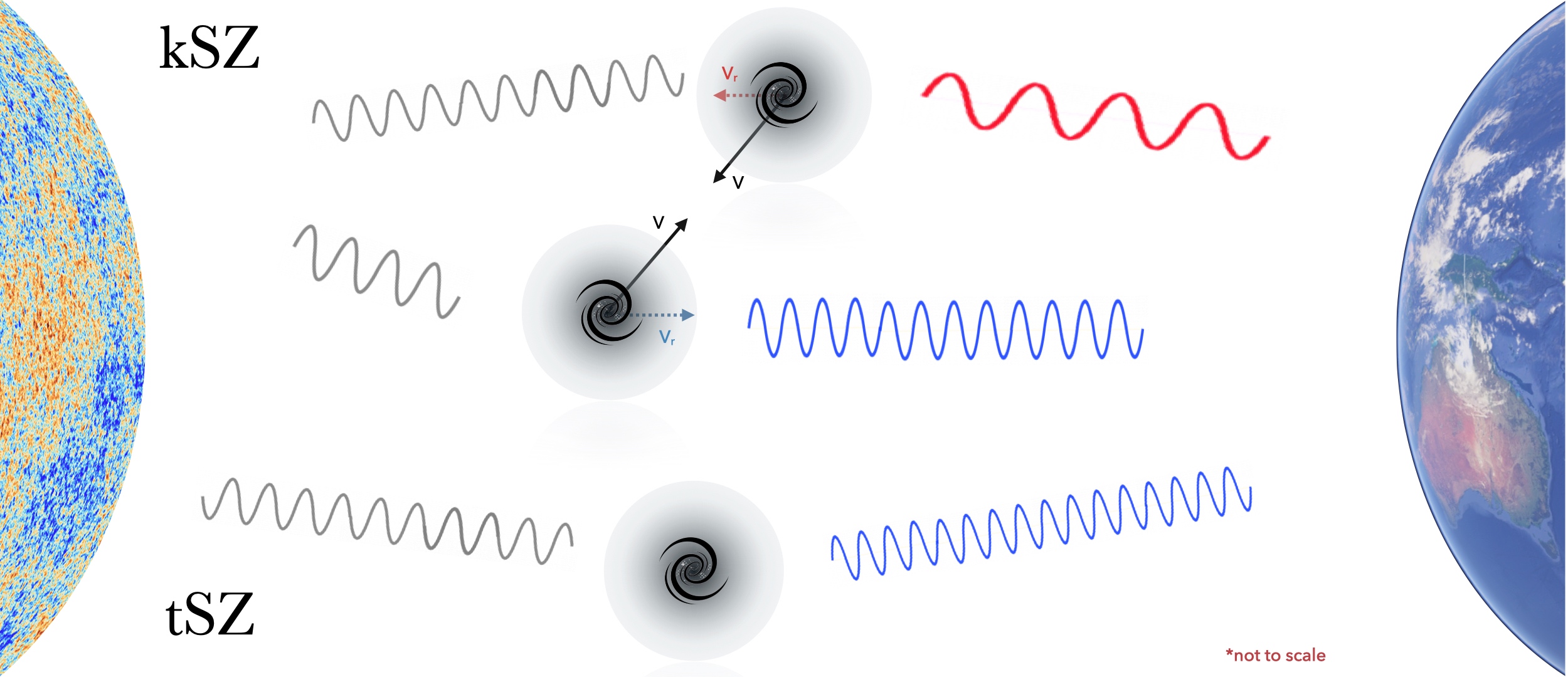}
    \caption{Illustration of the thermal Sunyaev-Zel’dovich (tSZ) and kinematic Sunyaev-Zel’dovich (kSZ) effects. The tSZ effect (bottom) occurs when CMB photons are scattered by hot electrons in galaxy clusters, leading to a frequency-dependent distortion in the CMB spectrum (blue-shifted photons). The kSZ effect (top) arises due to the bulk motion of galaxy clusters relative to the CMB rest frame, introducing a Doppler shift in the scattered CMB photons. The colors represent shifts in photon frequencies, with red indicating a positive radial velocity (receding cluster) and blue a negative radial velocity (approaching cluster) relative to the observer.}
    \label{fig:sz_effects}
\end{figure*}

One of the best-known and most extensively studied secondary anisotropies in the CMB is the SZ effect \citep{Sunyaev_1970, Sunyaev_1972}. The SZ effect originates from the inverse Compton scattering between CMB photons and free electrons in a hot, ionized gas along the line-of-sight, typically found in galaxy clusters (see Fig.~\ref{fig:sz_effects}).
The SZ effect is broadly classified into two components: the thermal SZ (tSZ) effect and the kinematic SZ (kSZ) effect. 
The tSZ effect arises from the scattering of CMB photons by the random thermal motions of electrons in the hot gas, resulting in a distinctive spectral distortion of the CMB. In contrast, the kSZ effect is caused by the bulk motion of electrons relative to the CMB rest frame, leading to a Doppler shift of the CMB photons without altering their overall Planckian spectrum.
These two effects can be considered as first- and second-order terms in the scattering of photons with a Planck distribution by moving electrons. The kSZ effect, being a first-order process, shifts the CMB spectrum without distorting its blackbody nature, whereas the tSZ effect, as a second-order process, induces spectral distortions. 
Typically, the thermal velocities of electrons within clusters are much larger than their bulk velocities, causing the tSZ effect to dominate. For typical cluster masses and velocities, the amplitude of the kSZ effect is approximately an order of magnitude smaller than that of the tSZ effect. For detailed reviews, see \citet{birkinshaw99} and \citet{carlstrom02}.

\subsubsection{Thermal SZ}
\label{sec:tsz}
The tSZ effect arises from the inverse Compton scattering of CMB photons by high-energy electrons in the hot intracluster medium of galaxy clusters. As CMB photons traverse this ionized gas, interactions with electrons at temperatures of $10^7$ to $10^8$ K upscatter the photons to higher energies. This process imprints a characteristic spectral distortion on the CMB, providing a powerful tool for studying galaxy clusters and the large-scale structure of the Universe.

The change in temperature/intensity of the CMB due to the tSZ effect can be described by the dimensionless Compton $y$-parameter, which quantifies the magnitude of the effect along a line-of-sight through the cluster. This change in temperature $\Delta T_{\rm tSZ}$ in the CMB is given by:
\begin{equation}
\label{eq:deltaT_tSZ}
    \frac{\Delta T_{\rm tSZ}}{T_{\rm CMB}} = f(x) y \, ,
\end{equation}
where $f(x)$ accounts for the frequency dependence of the tSZ, and the $y$-parameter is given by:
\begin{equation}
\label{eq:ycompt}
    y = \int \frac{k_B T_e}{m_e c^2} n_e \sigma_T \, \diff l \, ,
\end{equation}
where $k_B$ is the Boltzmann constant, $T_e$ is the electron temperature, $m_e$ is the electron mass, $c$ is the speed of light, $n_e$ is the electron number density, $\sigma_T$ is the Thomson cross-section, $\diff l$ is the differential path length along the line-of-sight.
This parameter essentially measures the integrated pressure of the electrons along the line-of-sight, making the tSZ effect sensitive to the electron distribution and thermal energy within the cluster.
The frequency dependence of the tSZ $f(x)$ is given as
\begin{equation}
\label{eq:fsz}
    f(x) = \left( x \frac{e^x + 1}{e^x - 1} - 4 \right)\, ,
\end{equation}
where $x = \frac{h \nu}{k_B T_{\mathrm{CMB}}}$. 
In general, observations in millimeter range are expressed in terms of specific intensity. For this purpose, the spectral distortion induced by the tSZ effect can be quantified by the change in the CMB intensity 
$\Delta I_{\rm tSZ}$, which depends on the frequency $\nu$ of the CMB photon. This change in intensity is described by:
\begin{equation}
    \Delta I_{\text{tSZ}}(\nu) = y \, I_0 \, g(x)\, ,
\end{equation}
where $I_0 = \frac{2 (k_B T_{\text{CMB}})^3}{(h c)^2}$ is the overall intensity scale, and the frequency dependence is given as 

\begin{equation}
    g(x) = \frac{x^4 e^x}{(e^x - 1)^2}\left( x \, \frac{e^x + 1}{e^x - 1} - 4 \right) \, .
\end{equation}

This frequency dependence produces a distinctive signature: at low frequencies (below 217 GHz), the tSZ effect manifests as a decrease in intensity, while at high frequencies (above 217 GHz), it appears as an increase in intensity. At approximately 217 GHz, the effect is null, as the energy gain and loss in the photon population cancel out, marking this as the crossover point (see Fig.~\ref{fig:tszkszcurves}). This characteristic frequency dependence enables a separation of the tSZ effect from other background radiation sources, such as synchrotron and dust emission.

The unique properties of the tSZ effect make it an invaluable tool for cosmology.
As highlighted in Eq.~\ref{eq:deltaT_tSZ} and \ref{eq:ycompt}, the temperature shift induced by the tSZ is proportional to the integrated pressure along the line of sight, making it unaffected by the inverse-square law that diminishes optical or X-ray brightness with distance.
This independence from redshift enables the detection of distant and faint galaxy clusters, providing a powerful means to map the LSS.
By analyzing the amplitude and spatial distribution of the $y$-parameter across numerous clusters, we can infer their mass distributions and investigate the thermal history of the universe, as well as energy feedback processes—such as those driven by active galactic nuclei (AGN)—that heat and redistribute gas within and around clusters

The advent of arcminute-scale resolution experiments such as SPT, ACT, and \textit{Planck} has transformed galaxy cluster science by enabling the identification of approximately mass-limited samples of massive galaxy clusters via the tSZ effect \citep[e.g.,][]{Staniszewski_2009,Menanteau_2010,reichardt13,Hasselfield_2013,planck15-27,bleem15b,Huang_2020,hiltonACT:2020,bleem20}. These tSZ-selected cluster catalogs have been instrumental in placing stringent constraints on cosmological models \citep[e.g.,][]{planck15-24,salvati17,zubeldia19,bocquet19}.
In addition, high-resolution, multi-frequency observations of the microwave sky have enabled the extraction of the thermal SZ Compton-$y$ parameter over vast regions using advanced component-separation algorithms \citep[e.g.,][]{planck15-22,madhavacheril19,Bleem_2022,Coulton_2024}. These measurements facilitate precise studies of the pressure profiles of galaxy clusters, the thermal state of the intracluster medium, and even the temperature of filaments connecting galaxy clusters within the cosmic web.
Looking ahead, future CMB surveys such as CMB-S4 promise to significantly expand our understanding of galaxy clusters.
With its wide survey covering $f_{\rm sky} = 50\%$, CMB-S4 is expected to detect approximately 75,000 clusters, while its deep survey covering $f_{\rm sky} = 3\%$ will detect around 14,000 clusters. 
Remarkably, about 1,350 of these clusters will be at $z \geq 2$, enabling unprecedented studies of cluster evolution at high redshifts \citep{Raghunathan_2022}.

\paragraph{Relativistic tSZ}
The relativistic tSZ effect is a generalization of the standard tSZ effect, incorporating corrections that arise due to the high thermal velocities of electrons in galaxy clusters. In the standard tSZ framework, the interaction between the CMB photons and the hot, ionized gas (intracluster medium) is treated under the assumption of a non-relativistic electron distribution, where the thermal electron velocities are much smaller than the speed of light. However, in massive clusters ($M \gtrsim 4\times 10^{14}M_{\odot}$) with very high temperatures—such as those exceeding $10$ keV—this approximation begins to break down, as the electrons exhibit relativistic velocities, i.e. $v_e^{\rm} = \sqrt{2k_B T_e/m_e} \gtrsim 0.1c$. These relativistic corrections modify both the amplitude and the frequency dependence of the tSZ spectral distortion  \citep[e.g.][]{Wright_1979, Rephaeli1995, Itoh1998}. The relativistic corrections are significant because they affect the spectral shape of the tSZ effect, altering the location and amplitude of the crossover frequency (where the tSZ effect transitions from a decrement to an increment relative to the CMB intensity). More specifically, Eq.~\ref{eq:fsz} modifies to 
\begin{equation}
    f(x) = \left( x \frac{e^x + 1}{e^x - 1} - 4 \right) \left(1 + \delta_{\rm rel}(x, T_e) \right)\, ,
\end{equation}
where $\delta_{\rm rel}(x, T_e)$ is the relativistic correction to the frequency dependence of tSZ.
These corrections are described by higher-order terms in the electron temperature $k_B T_e / m_e c^2$. 

The relativistic tSZ effect is computed using a more precise treatment of the photon-electron Compton scattering process, accounting for the full relativistic kinematics. This results in a spectral distortion given by a series expansion or more exact numerical integration. The primary impact of these corrections is at higher frequencies ($\nu > 200$ GHz), where they become essential for accurately modeling the tSZ signal in hot clusters.

Studying the relativistic tSZ effect provides valuable insights into the thermodynamic state of galaxy clusters. 
Since the magnitude and spectral features of the relativistic corrections depend on the cluster’s electron temperature, precise measurements of the tSZ effect across a wide frequency range allow us to probe the temperature distribution of the intracluster medium \citep[e.g.,][]{Hurier_2016,Erler2018,Coulton_2024, Remazeilles2024}. 
This is especially useful for distinguishing between thermal and non-thermal components, as well as for testing cosmological models through the relationship between the cluster gas and total mass.

\subsubsection{Kinematic SZ}
The kSZ effect is caused by the Doppler shift of CMB photons as they scatter off free electrons in the intracluster medium (ICM) of galaxy clusters. Unlike tSZ effect, which arises from the thermal motion of electrons, the kSZ effect results from the bulk peculiar velocity of the cluster relative to the CMB rest frame. This effect introduces a shift in the observed CMB temperature without altering its spectral shape.

The kSZ effect produces a temperature fluctuation $\Delta T_{\rm kSZ}$ in the CMB, which is given by:
\begin{equation}
    \frac{\Delta T_{\rm kSZ}}{T_{\rm CMB}} = - \int n_e e^{-\tau_e} \sigma_T \frac{{\bf v} \cdot \nver}{c} dl \, ,
\end{equation}
where ${\bf v}$ is the peculiar velocity of the cluster and $\nver$ is the unit vector pointing along the line-of-sight. The term ${\bf v} \cdot \nver$ represents the component of the cluster's peculiar velocity along the observer's line-of-sight ($v_r$, or the radial velocity). $\tau_e$ is the optical depth to Thomson scattering, and the integral accounts for the electron density $n_e$ along the line-of-sight, weighted by the velocity. 

\begin{figure*}
    \centering
    \includegraphics[width=0.8\textwidth]{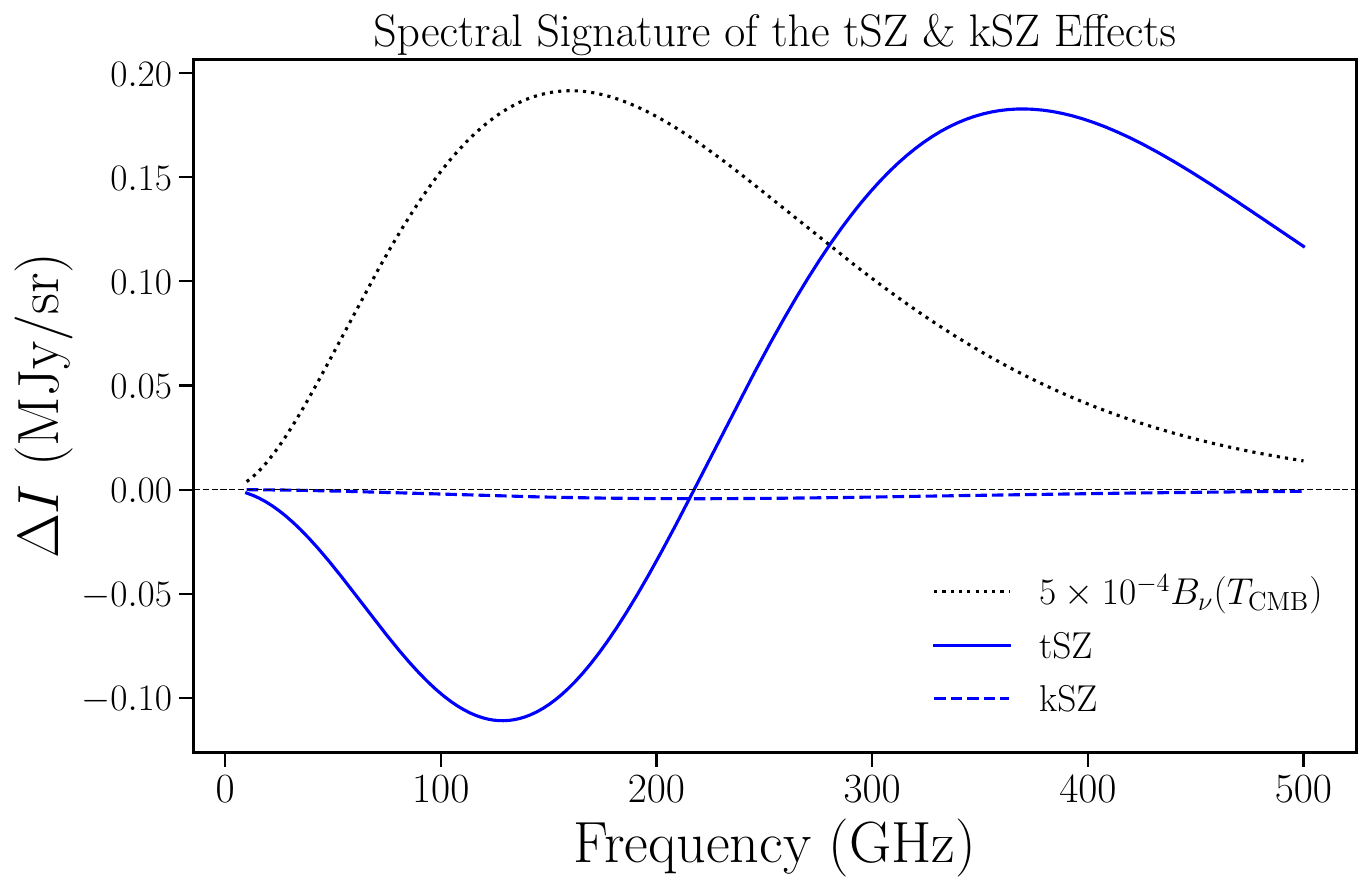}
    \caption{Frequency-dependent changes in CMB intensity caused by the tSZ (solid blue) and kSZ (dashed blue) effects. The dotted curve represents the thermal CMB spectrum at a temperature of 2.7 K, scaled by a factor of $5 \times 10^{-4}$. The kSZ effect preserves the blackbody nature of the CMB spectrum, while the tSZ effect introduces a spectral distortion with a null point around 217 GHz. These curves were generated using a Compton parameter $y = 10^{-4}$, a cluster peculiar velocity of $1000$ km/s, and an optical depth $\tau_e = 10^{-3}$.}
    \label{fig:tszkszcurves}
\end{figure*}

\noindent The optical depth, $\tau_e$, of a cluster is given by:
\begin{equation}
\label{eq:taue}
    \tau_e = \int n_e \sigma_T dl \, ,
\end{equation}
and the kSZ temperature fluctuation simplifies to:
\begin{equation}
    \frac{\Delta T_{\rm kSZ}}{T_{\rm CMB}} = - \tau_e \frac{v_r}{c} \, .
\end{equation}
Unlike the tSZ effect, the kSZ effect preserves the blackbody shape of the CMB spectrum. This means it results in a uniform shift in intensity at all frequencies proportional to the peculiar velocity of the cluster (see Fig.~\ref{fig:tszkszcurves}). 
The kSZ signal is challenging to detect because it lacks the distinct spectral signature of the tSZ effect and is typically smaller in magnitude.

The kSZ effect provides unique insights into the large-scale velocity field of the universe \citep[e.g.,][]{Deutsch_2018, Smith_2018}. 
By measuring the kSZ signal, we can estimate the peculiar velocities of galaxy clusters, revealing information about the distribution of matter and the LSS dynamics. 
This is particularly useful for testing cosmological models, including the influence of dark energy and modified gravity on the growth of structure \citep[e.g.][]{DeDeo_2005, bhattacharya07, mueller15, ma13, Alonso_2016,bianchini16}, as well as primordial non-Gausianities \citep{M_nchmeyer_2019}.
Furthermore, kSZ measurements can constrain the baryon content of the universe, as the signal depends on the electron density along the line-of-sight. 
Combined with weak gravitational lensing and tSZ measurements, the kSZ effect allows for a detailed study of the connection between baryonic and dark matter in galaxy clusters \citep[e.g.,][]{Battaglia:2017neq,Schaan_2021, Amodeo_2021}.

Detecting the kSZ effect is challenging due to its small amplitude, contamination from primary CMB anisotropies, and other astrophysical foregrounds. 
As a result, individual clusters' velocities are currently measured from mm-wave observations for only a small number of objects \citep[e.g.,][]{sayers13,adam17}. 
Since the kSZ effect preserves the blackbody spectrum of the primary CMB, it cannot be isolated using standard component separation techniques alone. 
Instead, the signal is detected by combining CMB data with LSS observations through statistical methods or estimators.
Early detections of the kSZ effect were achieved by applying the pairwise momentum method to ACT data combined with BOSS spectroscopic galaxies in 2012 \citep{hand12}, followed by additional detections using \textit{Planck}, SPT, and ACT data in conjunction with spectroscopic and photometric LSS samples \citep{debernardis17,soergel16,calafut21,Schiappucci_2023}. 
Velocity-weighted stacking techniques have been employed to constrain baryon density profiles, as demonstrated with ACT data and BOSS galaxies \citep{Schaan:2015,Schaan_2021,Amodeo_2021}, and more recently with DESI photometric LRGs \citep{hadzhiyska2024evidencelargebaryonicfeedback}. 
The first detection using velocity reconstruction was recently achieved with ACT data and DESI LRGs \citep{mccarthy2024atacamacosmologytelescopelargescale}.
Other classes of estimators that are sensitive to the $\langle TTg \rangle$ bispectrum (where `$g$' represents the galaxy over-density field), also known as projected-fields estimators, involve filtering and squaring CMB maps before cross-correlating with LSS, represent distinct approaches to extracting kSZ signals \citep[e.g.,][]{Hill_2016,Kusiak_2021}.

Another significant contribution to the kSZ signal originates from the epoch of reionization, where local variations in the ionized fraction and free electron density create arcminute-scale temperature anisotropies in the CMB. This effect is commonly referred to as the ``patchy" kSZ effect \citep[e.g.,][]{mcquinn05,shaw12,battaglia13,Park_2013}.
Distinguishing the patchy kSZ signal from the low-redshift kSZ contribution is challenging. \citet{smith17} proposed using the kSZ trispectrum as a method to isolate the reionization-era contribution. However, recent analyses applying this approach to SPT/\textit{Herschel} and ACT data have so far only yielded upper limits on the kSZ trispectrum. These studies found that foreground contamination poses a significant obstacle, with foreground signals dominating over the expected kSZ signal \citep{Raghunathan2024,MacCrann:2024ahs}.

\paragraph{Rotational kSZ}
The rotational kSZ effect is a variant of the kSZ effect that arises from the rotational motion of galaxy clusters. Unlike the standard kSZ effect, which is caused by the bulk peculiar motion of the cluster along the line-of-sight, the rotational kSZ effect originates from the coherent tangential motion of intracluster gas around the center of mass of the cluster. This rotational motion creates a dipole-like pattern in the CMB temperature fluctuations across the cluster, with opposite sides of the cluster showing slightly redshifted and blueshifted signals, respectively, due to the Doppler effect. The amplitude of the rotational kSZ effect is proportional to the rotational velocity of the gas and the optical depth ($\tau_e$). Observing this effect provides a unique way to measure angular momentum in galaxy clusters and to study the dynamics of gas in cluster halos, offering insights into the processes driving the assembly and evolution of large-scale structure. Detecting the rotational kSZ effect, however, is observationally challenging due to its weak signal and the need to separate it from other kSZ contributions and astrophysical foregrounds.
For example, \citet{Baxter_2019} analyzed \textit{Planck} data using 13 galaxy clusters from SDSS-DR10, reporting approximately $2\sigma$ evidence of a rotational kSZ signal with an amplitude and morphology consistent with theoretical expectations.

\subsection{Polarized SZ}
While SZ effects predominantly influence the intensity of the CMB, polarization can also occur under specific circumstances. The main sources of SZ polarization are quadrupole anisotropies in the incoming CMB radiation \citep{Sazonov_1999} and the peculiar motion of galaxy clusters \citep{Sunyaev_1980, Sazonov_1999}. Quadrupole-induced polarization occurs because the Thomson scattering process imprints the anisotropy of the incoming radiation onto the scattered light. This effect depends on the local CMB quadrupole observed at the location of the scattering electrons and is the dominant source of polarization in SZ. It is typically expected to be on the order of micro-Kelvin or smaller. Detection of this effect could allow measurements of the optical depth $\tau_e$ of the cluster as it is proportional to $\tau$ times the tSZ effect. In principle, it could thus separate $T_e$ and $\tau_e$ by measuring tSZ (Eq.~\ref{eq:ycompt} and \ref{eq:taue}). 

In contrast, motion-induced polarization arises from the relative velocity between the cluster and the observer. This effect is proportional to $(v_t/c)^2 \tau_e$, where $v_t$ is transverse velocity of the cluster. Typically this effect is of the order of nano-Kelvin and is sensitive to the transverse velocity component of the cluster with respect to our line-of-sight, which is difficult to measure otherwise. The resulting polarization vector is perpendicular to the plane formed by the line-of-sight and the velocity vector.

Two other sources of polarization in SZ are double scattering-induced polarization, and Faraday rotation-induced polarization due to presence of magnetic fields in clusters \citep[e.g.][]{Aghanim_2008}. These effects are expected to be smaller than the dominant quadrupole-induced polarization. 

Polarization of the SZ effect is typically much weaker than the intensity signals but carries unique information. By detecting and analyzing the polarized SZ signals, cosmologists can probe the dynamics of galaxy clusters, map large-scale velocity fields, and potentially constrain the evolution of the CMB quadrupole, offering insights into early universe physics and large-scale structure evolution. 

\subsection{Patchy screening}

\begin{figure}
\centering
\begin{subfigure}{.5\textwidth}
  \centering
  \includegraphics[width=.9\linewidth, height=7cm]{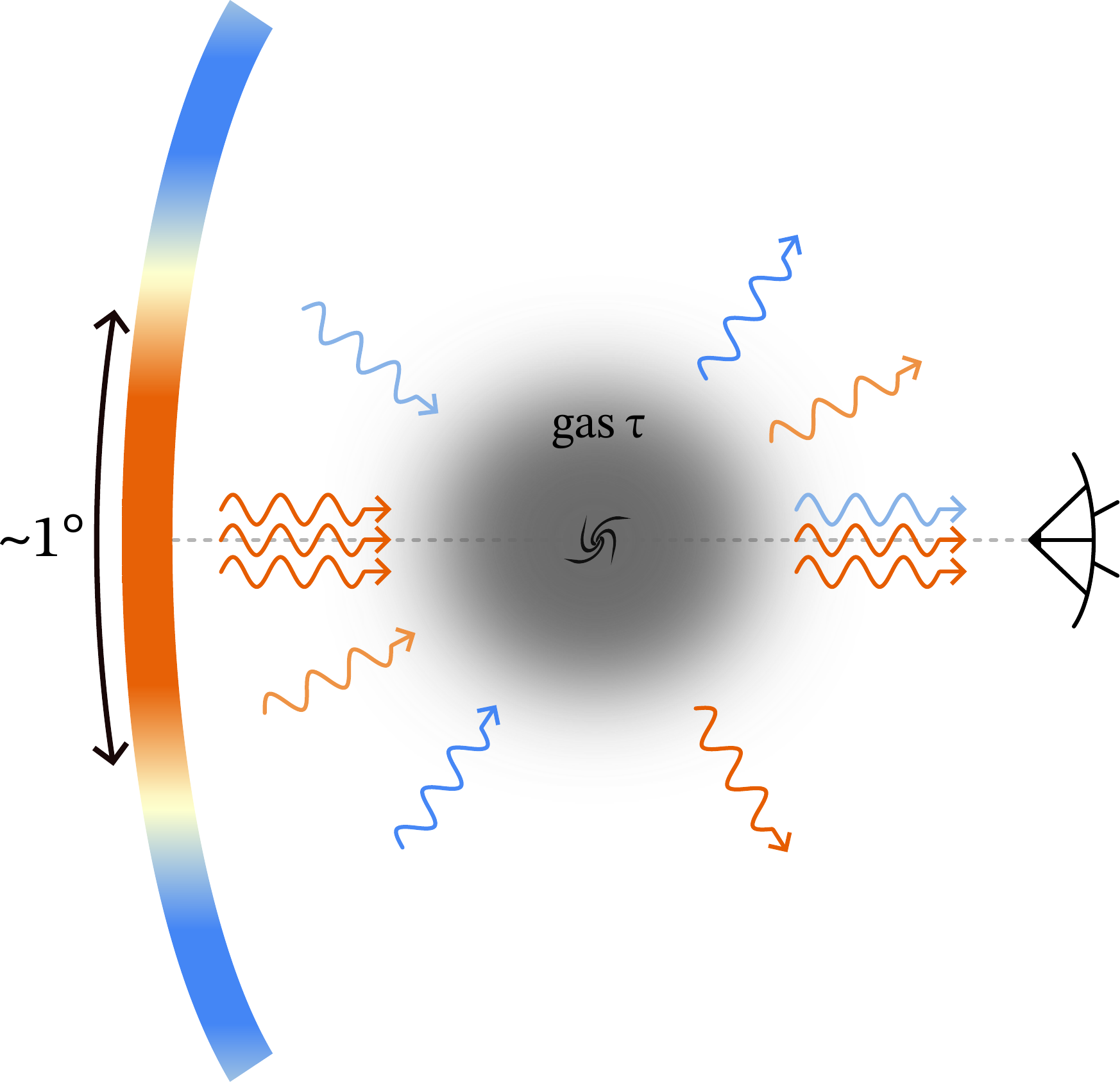}
\end{subfigure}%
\begin{subfigure}{.5\textwidth}
  \centering
  \includegraphics[width=.9\linewidth, height=7cm]{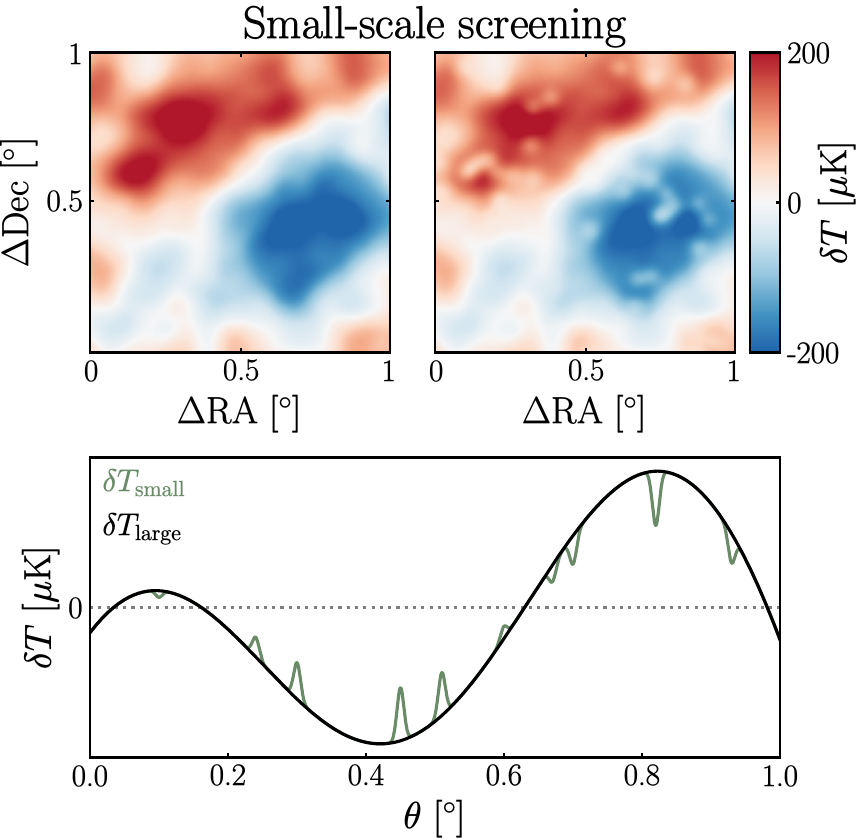}
\end{subfigure}
\caption{{\bf \textit{Left}}: Intuitive understanding of the patchy screening effect. We have photons coming from typical 1 deg CMB fluctuations from left that get Thomson scattered by electrons in gas around a galaxy in the line-of-sight (LOS). As a result, some of these photons are scattered out of the LOS and are replaced by photons from all other directions scattered into our LOS with the temperature averaged across all directions. This has an effect of `smoothing' CMB anisotropies in a given LOS. Credit: Theo Schutt. {\bf \textit{Right}}: Top- 1 deg$^2$ simulated map of a CMB realization
unscreened (left) and screened by 5’ FWHM 2D Gaussian $\tau$
profiles placed at a fixed realization of galaxy positions in the foreground (right). The $\tau$ amplitude is greatly exaggerated for visual effect. Bottom- A 1D schematic illustrating the small angular scale temperature deviations $\delta T_{\rm small}$ induced by the patchy screening effect (green). Unlike any other foreground, these deviations are (anti-)correlated with the large-scale primary
CMB anisotropies $\delta T_{\rm large}$ (black). This effectively `smooths' CMB anisotropies bringing the small-scale anisotropies close to the mean of the map as can be seen by comparing the unscreened and screened simulated maps. Plot taken from \citet{Schutt_2024}.} 
\label{fig:patchytau}
\end{figure}

Apart from the inverse Compton scattering which gives rise to the SZ effects, CMB photons also undergo Thomson scattering by electrons in gas around galaxies and in clusters. This scattering damps the primary CMB anisotropies and has two components. The isotropic component of this scattering scales with the optical depth to reionization which is one of the six standard parameters of the $\Lambda$CDM cosmology \citep{Planck_2020}. 
This relationship connects the observed CMB anisotropies to the intrinsic ones through:
\begin{equation}
\label{eq:patchytau}
    \Delta T(\nver) = \Delta T^0(\nver)e^{-\tau(\nver)} \, ,
\end{equation}
where $\Delta T(\nver)$ and $\Delta T^0(\nver)$ are the observed and intrinsic CMB temperature anisotropy in a given direction $\nver$ respectively, and $\tau(\nver)$ is the optical depth to Thomson scattering in that patch of the sky.  
However, the distribution of electrons responsible for this scattering is anisotropic, meaning that some lines of sight contain more electrons than others.
This leads to anisotropic damping, which spatially modulates the underlying CMB anisotropies and introduces a new source of small-scale $\leq 1^{\circ}$ anisotropies in the CMB maps. 
This phenomenon is known as ``patchy screening" of the CMB photons. 

Patchy screening arises from two distinct phases: (1) during the epoch of reionization, when large spatial fluctuations in the ionization fraction were present, and (2) in the late-time Universe  ($z < 5$) when gravitationally collapsed objects have large electron overdensities. 
Consequently, patchy screening traces anisotropies in the optical depth to Thomson scattering, as ionized or collapsed regions are unevenly distributed across the sky. This can be expressed as: $\tau(\nver) = \overline{\tau} + \delta \tau(\nver)$; where $\overline{\tau}$ and $\delta \tau(\nver)$ are the mean and fluctuations in $\tau$ respectively. Using this in Eq.~\ref{eq:patchytau} with $\delta \tau(\nver) \ll 1$, and absorbing $\overline{\tau}$ in the intrinsic CMB anisotropies gives
\begin{equation}
\label{eq:patchytau2}
    \Delta T(\nver) = -\delta \tau(\nver) \: \Delta T^0(\nver) \, .
\end{equation}
This expression is analogous to Eq.~\ref{eq:lens_firs_ord_exp} for weak lensing of the CMB. In the Fourier domain, patchy screening, like CMB lensing, couples different modes of the CMB. These mode couplings form the basis for constructing quadratic estimators of the screening field \citep{Dvorkin_2009}. 
While Eqs.~\ref{eq:patchytau} and \ref{eq:patchytau2} are written for CMB temperature maps, they also apply to the $Q$ and $U$ polarization maps. Similar to CMB lensing, patchy screening generates a $B$-mode polarization signal. 

A clear separation of scales emerges in Eq.~\ref{eq:patchytau2}, when considering the effect of screening on small scales. Denoting small and large scales with subscripts S and L respectively, Eq.~\ref{eq:patchytau2} becomes 
\begin{equation}
\label{eq:patchytau3}
    \Delta T_S(\nver) = -\delta \tau_S(\nver) \: \Delta T_L^0(\nver) \, ,
\end{equation}
indicating that the screening effect on small scales is almost exclusively sourced by  configurations where small-scale variations in the optical depth modulate the large-scale CMB anisotropies. 
As a result of this separation of scales, \citet{Schutt_2024} were able to show that both the screened and unscreened temperature fluctuations can be measured from the same map, and that can give an estimate of patchy screening on small scales. 
This effect is illustrated in the bottom-right panel of Fig.~\ref{fig:patchytau}, which shows how small-scale CMB anisotropies due to screening arise from the modulation of large-scale CMB anisotropies by small-scale screening. 
Additionally, as seen from Eqs.~\ref{eq:patchytau2} and \ref{eq:patchytau3}, the sign of the screening anisotropy is always opposite to the sign of the corresponding large-scale CMB anisotropy in the same patch of sky.
This implies that screening effectively ``smooths" the CMB anisotropies in a given region, bringing them closer to the mean. This smoothing effect can be observed in the top-right panel of Fig.~\ref{fig:patchytau}, where the left and right simulated maps illustrate the unscreened and screened CMB anisotropies, respectively.

Finally, since patchy screening receives contributions from the epoch of reionization, it provides a valuable tool for studying the mechanisms of reionization. 
The patchy screening signal is sensitive to the morphology and timing of reionization, including the size and distribution of ionized bubbles, the nature of ionizing sources, and the properties of the surrounding IGM.
By analyzing this effect, we can infer key aspects of the history of reionization, such as its onset and the timescale over which it progressed \citep[e.g.,][]{Roy_2018, Namikawa_2021,BICEPKeck:2022kci, Bianchini_2023}.
In addition, since the damping caused by screening is proportional to the free electrons overdensity, the low redshift contribution to screening is an excellent probe of the gas in the circumgalactic medium \citep[e.g.][]{Roy_2023, Schutt_2024, Coulton_2024a}. 
The amplitude and angular scale of the patchy screening signal directly trace the projected gas profile, making screening a complementary probe to the kSZ effect, which also studies gas around galaxies. 
Similar to the tSZ effect, patchy screening is independent of redshift, making it a powerful tool for probing gas properties across a wide range of cosmic epochs, from low to high redshifts.


\section{Concluding Remarks}

In this chapter, we have explored the physical mechanisms behind secondary anisotropies in the CMB and reviewed their current observational status and future prospects. 
Secondary anisotropies arise from two broad classes of effects, categorized by their underlying sourcing mechanisms: gravitational and scattering. 
As CMB photons journey from the surface of last scattering to us, they interact with the LSS of the universe, providing a unique backlight that enables the imaging of both gravitational potentials and the gas distribution in the universe.
Gravitational effects such as weak gravitational lensing, the ISW effect, and the RS and ML effects allow us to probe the distribution and evolution of matter and gravitational potentials. 
Scattering effects, such as those caused by Thomson and inverse Compton scattering, include the tSZ and kSZ effects, as well as patchy screening, which provide valuable insights into the distribution, temperature, and motion of ionized gas in the universe.

This is an exciting time for CMB research, as the current generation of surveys—including \textit{Planck} \citep{planck18-6}, SPT \citep{reichardt20}, and ACT \citep{aiola20}—have begun to unlock the vast potential of secondary CMB anisotropies. 
These experiments have demonstrated the power of CMB maps to shed light on both cosmological and astrophysical phenomena, laying the foundation for the next generation of surveys.
Upcoming experiments such as the Simons Observatory \citep{Ade2019}, FYST/CCAT-prime \citep{ccat}, and CMB-S4 \citep{CMB-S4_22} promise to deliver transformative advancements. With their higher sensitivity, improved resolution, and broader frequency coverage, these surveys will provide a high-fidelity view of secondary anisotropies in both intensity and polarization. This unprecedented precision will enable us to probe fundamental physics and astrophysics in ways previously unattainable.

One particularly exciting frontier is the synergy between CMB and LSS surveys \citep{baxter2022snowmass2021opportunitiescrosssurveyanalyses}. 
The combination of these complementary probes will enable robust cross-correlation analyses, yielding tighter and more robust constraints on key questions in both cosmology and astrophysics. 
For example, these analyses will improve our understanding of the equation of state of dark energy, primordial non-Gaussianity, and the physics of galaxy biasing, intrinsic alignments, and baryonic feedback processes. 
The expected science return from these cross-correlation studies is multi-faceted, offering insights into both fundamental physics and the interplay of matter and radiation in the universe.
To fully capitalize on these opportunities, significant efforts are required to develop improved theoretical models and correlated simulations that can match the precision of these upcoming datasets.
Work in this direction has already begun \citep[e.g.,][]{Stein_2020,Han_2021,Omori2024,bayer2024halfdomemultisurveycosmologicalsimulations}, but continued advancements will be essential for ensuring robust and unbiased analyses.

In summary, the study of secondary CMB anisotropies represents a rapidly evolving field that bridges cosmology and astrophysics. 
Current and next-generation experiments, along with advances in theory and simulations, will further solidify the role of CMB secondary anisotropies as a cornerstone of precision cosmology and a window into the underlying structure and dynamics of the universe.

\begin{ack}[Acknowledgments]\\
We dedicate this chapter to the memory of Eric Baxter, whose contributions to this field and to the study of secondary anisotropies were both profound and inspiring. Eric was a brilliant researcher and a wonderful colleague, and his presence in our community is deeply missed. 
We also thank Yuuki Omori, Emmanuel Schaan, Theo Schutt, and Zhuoqi Zhang for their valuable feedback and for providing material that greatly enhanced this work.
FB acknowledges support by the Department of Energy, Contract DE-AC02-76SF00515.
\end{ack}


\bibliographystyle{Harvard}
\bibliography{reference}

\end{document}